\pdfoutput=1
\PassOptionsToPackage{pdfpagelabels=false}{hyperref}
\documentclass[fleqn,usenatbib,usedcolumn]{mnras}
\usepackage[british]{babel}             
\usepackage{newtxtext,newtxmath}
\usepackage[T1]{fontenc}

\usepackage{graphicx}    
\usepackage{amsmath}    
\usepackage{amssymb}    

\newcommand{\feh}{\ensuremath{[\textrm{Fe}/\textrm{H}]}}
\newcommand{\mh}{\ensuremath{[\textrm{M}/\textrm{H}]}}
\newcommand{\ebv}{\ensuremath{\textrm{E}(B-V)}}
\newcommand{\ejk}{\ensuremath{\textrm{E}(J-K_S)}}
\newcommand{\teff}{\ensuremath{\textrm{T}_\textrm{eff}}}
\newcommand{\logg}{\ensuremath{\log \textrm{g}}}
\newcommand{\gaia}{\textit{Gaia}~1}

\title[Confirmation of \gaia]{\textit{Siriusly}, a newly identified intermediate-age Milky Way stellar cluster: A spectroscopic study of \gaia}

\author[J. D. Simpson et al.]{J. D. Simpson${^{1,2,3}}$\thanks{email: jeffrey.simpson@aao.gov.au},  G. M. De Silva${^{1,4}}$, S. L. Martell$^{5}$, D. B. Zucker$^{1,2,3}$, \newauthor{A. M. N. Ferguson$^{6}$, E. J. Bernard$^{7}$, M. Irwin$^{8}$, J. Penarrubia$^{6}$, E. Tolstoy$^{9}$}\\
$^1$Australian Astronomical Observatory, 105 Delhi Rd, North Ryde, NSW 2113, Australia\\
$^2$Department of Physics and Astronomy, Macquarie University, Sydney NSW 2109, Australia\\
$^3$Research Centre in Astronomy, Astrophysics and Astrophotonics, Macquarie University, Sydney NSW 2109, Australia\\
$^4$Sydney Institute for Astronomy, School of Physics, A28, The University of Sydney, Sydney NSW 2006,  Australia\\
$^5$School of Physics, University of New South Wales, Sydney NSW 2052, Australia\\
$^{6}$Institute for Astronomy, University of Edinburgh, Blackford Hill, Edinburgh EH9 3HJ, UK\\
$^{7}$Universit\'e C\^ote d'Azur, OCA, CNRS, Lagrange, France\\
$^{8}$Institute of Astronomy, University of Cambridge, Madingley Road, Cambridge, CB3 0HA, UK\\
$^{9}$Kapteyn Astronomical Institute, University of Groningen, Landleven 12, 9747AD Groningen, Netherlands\\}

\date{Accepted 2017 July 23. Received 2017 July 23; in original form 2017 March 15}

\pubyear{2017}

\begin{document}
\label{firstpage}
\pagerange{\pageref{firstpage}--\pageref{lastpage}}
\maketitle

\begin{abstract}
We confirm the reality of the recently discovered Milky Way stellar cluster \gaia\ using spectra acquired with the HERMES and AAOmega spectrographs of the Anglo-Australian Telescope. This cluster had been previously undiscovered due to its close angular proximity to Sirius, the brightest star in the sky at visual wavelengths. Our observations identified 41 cluster members, and yielded an overall metallicity of $\feh=-0.13\pm0.13$ and barycentric radial velocity of $v_r=58.30\pm0.22$~km/s. These kinematics provide a dynamical mass estimate of $12.9^{+4.6}_{-3.9}\times10^3$~M$_{\sun}$. Isochrone fits to \textit{Gaia}, 2MASS, and Pan-STARRS1 photometry indicate that \gaia\ is an intermediate age ($\sim3$~Gyr) stellar cluster. Combining the spatial and kinematic data we calculate \gaia\ has a circular orbit with a radius of about 12~kpc, but with a large out of plane motion: $z_\textrm{max}=1.1^{+0.4}_{-0.3}$~kpc. Clusters with such orbits are unlikely to survive long due to the number of plane passages they would experience.
\end{abstract}




\section{Introduction}\label{sec:intro}

The ESA \textit{Gaia} mission has the goal of constructing the largest and most precise 6D space catalogue ever made. It is measuring the positions, distances, space motions and many physical characteristics of some one billion stars in our Galaxy and beyond \citep{GaiaCollaboration2016}. The first data release \citep{Brown2016} has already been used to identify previously unknown co-moving pairs of stars \citep{Oh2016,Andrews2017}, determine parallactic distances to globular clusters \citep{Watkins2016}, measure the proper motion of the distant globular cluster NGC~2419 by combining \textit{Gaia} and \textit{HST} data \citep{Massari2017}, and show that M40 is in fact just two unrelated stars and not a true binary \citep{Merrifield2016}.

\citet{Koposov2017} took advantage of a number of unique capabilities of \textit{Gaia} (i.e., no weather and sky brightness variations; low-to-no spurious detections; excellent star/galaxy discrimination) to search for stellar overdensities using a modified method they had previously applied with great success to ground-based surveys \citep[e.g.,][]{Koposov2007,Koposov2015}. Their search of the \textit{Gaia} catalogue identified 259 candidates overdensities, of which 244 had clear associations with previously known clusters and dwarf galaxies. Of the unknown overdensities, two were statistically significant enough to warrant quick publication: \gaia\ and \textit{Gaia}~2. Of particular note is \gaia\ which is located only 11~arcmin from Sirius (though their physical separation is $\sim4$~kpc), the brightest star in the night sky. This cluster likely would have been previously identified had it not been for this proximity, which has hidden its existence from astronomers.

Beyond the novelty of being previously undiscovered, the cluster parameters estimated by \citet{Koposov2017} --- 6~Gyr, 14000~M$_\textrm{\sun}$, $\feh=-0.7$ --- suggest that \gaia\ is an interesting target in its own right due to its being on the border between open and globular clusters. It is about 1~kpc out of the plane of the Galaxy, which might suggest it is an open cluster. The photometry shows a populated red clump region, indicative of an intermediate age, metal-rich cluster \citep[see the review by][and references therein]{Girardi2016}. However, the available photometry can only provide so much information, with a complete picture of the cluster's chemistry and kinematic only possible when the photometry is combined with spectroscopy.

It is important to spectroscopically observe purported clusters to confirm that the stars truly have similar kinematics and chemistry. As an example, Lod\'{e}n~1 \citep{Loden1980} had been catalogued as a 2-Gyr stellar cluster at a heliocentric distance of 360~pc in the Database for Galactic Open Clusters \citep[WEBDA,][]{Mermilliod2003}. Such properties would make it a very useful old, nearby cluster target for calibration and benchmarking of large stellar surveys. But when \citet{Han2016} investigated the photometry and kinematics of the cluster, they determined that Lod\'{e}n~1 was ``neither old, nor nearby, nor a cluster!'' The positional and kinematic information of \textit{Gaia}, combined with the chemical information of large stellar surveys (e.g., GALAH, APOGEE, Gaia-ESO, 4MOST, WEAVE) will likely lead to the ``de-identification'' of several other putative clusters.

In this paper we confirm that \gaia\ is a kinematically distinct cluster of stars, using spectra acquired with the Anglo-Australian Telescope HERMES and AAOmega spectrographs. This paper is structured as follows: Section \ref{sec:observations} details the observations; Section \ref{sec:rv_feh} explains how the radial velocities were determined for the stars; Section \ref{sec:feh} discusses the metallicities estimated from the high-resolution HERMES spectra, and estimated from the CaT lines in the AAOmega spectra; Section \ref{sec:members} combines the radial velocity and metallicity information to identify members; and Section \ref{sec:parameters} discusses the results and what they mean for the overall cluster properties and its orbit.

\section{Observations}\label{sec:observations}
\gaia\ was observed with two of the spectrographs of the 3.9-metre Anglo-Australian Telescope over three nights: on the night of 2017 February 15 with the four-armed high-resolution HERMES spectrograph; and on the nights of 2017 February 24 \& 26 with the two-armed AAOmega spectrograph. For all the observations, the light was fed to the instruments using the 392-fibre Two-Degree Field (2dF) optical fibre positioner top-end \citep{Lewis2002}.

HERMES simultaneously acquires spectra using four independent cameras with non-contiguous wavelength coverage totalling $\sim1000$~\AA\ at a spectral resolution of $R\approx28,000$ \citep{Sheinis2015}. Its fixed wavelength bands were chosen primarily for the on-going GALAH survey \citep[blue: 4715--4900~\AA; green: 5649--5873~\AA; red: 6478--6737~\AA; near-infrared: 7585--7887~\AA;][]{DeSilva2015a,Martell2017}. 

AAOmega simultaneously acquires spectra using independent blue and red cameras \citep{Sharp2006}. For these observations, the blue camera was fitted with the 580V grating ($R\sim$1200; 3700--5800~\AA) and the red camera fitted with the 1700D grating (R$\sim10000$; 8340--8840\AA). The latter grating is designed for observations of the near-infrared calcium triplet lines around 8600\AA, which allows for metallicity estimation and precise radial velocity measurement.

\gaia\ is located at $\textrm{RA}=6^\textrm{h}45^\textrm{m}53^\textrm{s},\textrm{Dec}=-16\degr45\arcmin00\arcsec$ and has an angular extent of $\sim15$~arcmin. As mentioned in Section \ref{sec:intro}, the centre of \gaia\ is located only $11\arcmin$ from the $V=-1.5$ Sirius system. This meant we had two main concerns when observing \gaia: that diffraction spikes from Sirius could coincide with fibres; and that the scattered light from Sirius could be so large as to overwhelm the light from the target stars which are 13--18 magnitudes fainter. The extent and brightness of the diffraction spikes from Sirius were difficult to predict, so our primary mitigation method was to simply avoid placing fibres within 10~arcmin of Sirius. To reduce the scattered light from Sirius we used the 2dF plate that was coated black, and the field was centred on Sirius, with the cluster off-centre. Placing the cluster off-centre does have the trade-off of reducing the number of targets that can be observed, as the fibres of 2dF have a maximum allowed offset from their radial positions, and cannot be placed across the centre of the plate. 

The observed targets were selected from a catalogue created by cross-matching \textit{Gaia} DR1 with The Two Micron All Sky Survey (2MASS) catalogue \citep{Skrutskie2006}. This cross-match was performed using the \textit{Gaia} archive and joining their \texttt{gaia\_source}, \texttt{tmass\_best\_neighour} and \texttt{tmass\_original\_valid} tables on \texttt{source\_id} and \texttt{tmass\_oid}. For the target identification within the Panoramic Survey Telescope and Rapid Response System Data Release 1 \citep[Pan-STARRS1 or PS1;][]{Chambers2016,Magnier2016a}, a closest neighbour positional search was used between the \textit{Gaia}-2MASS table and a table created from PS1's \texttt{ObjectThin} and \texttt{MeanObject} tables, with a minimum of 11 PS1 epoch detections. We found that 11 detections of an object was the minimum required to remove the large number of spurious objects in PS1 associated with the diffraction pattern of Sirius in the PS1 images. For the 1~degree around Sirius, there were a median of 56 detections per target in our final catalogue.

\begin{figure*}
    \includegraphics[width=\textwidth]{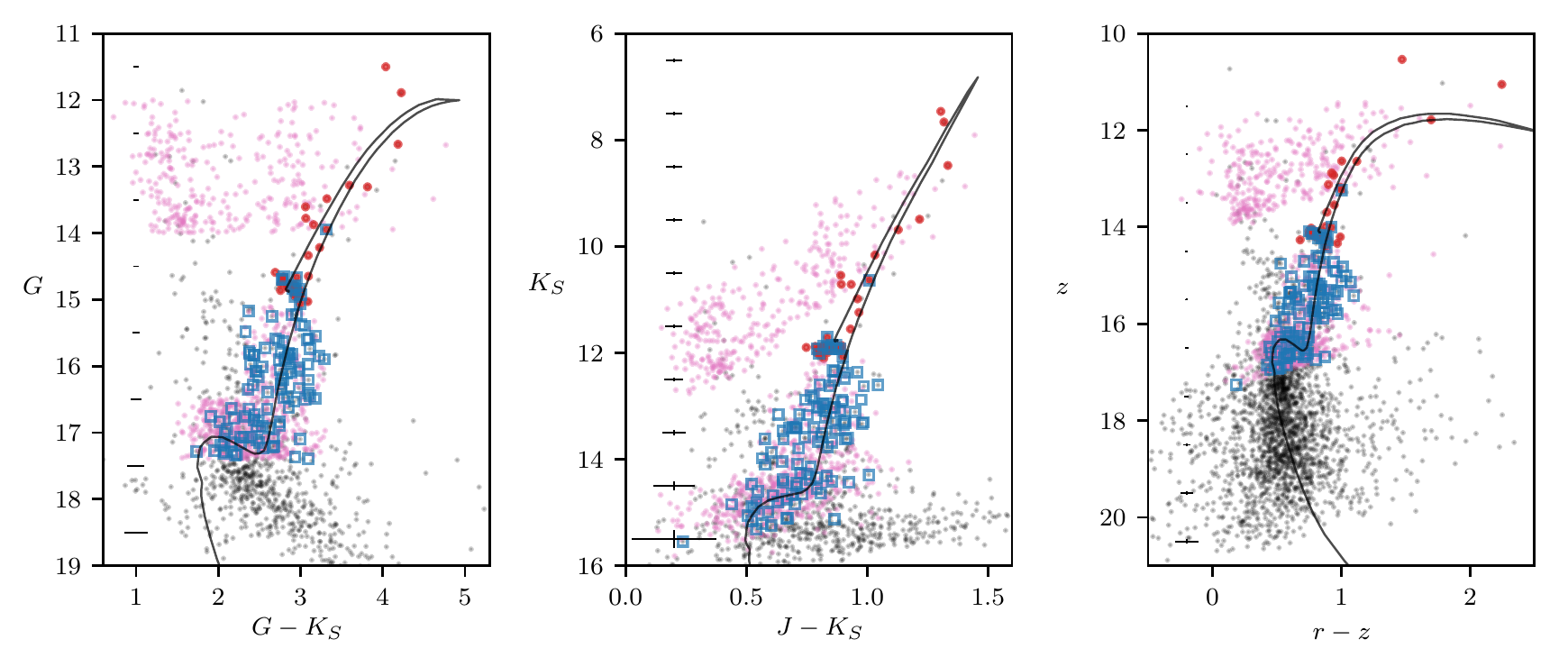}
    \caption{Colour-magnitude diagrams of \gaia\ in the various photometric catalogues. The left panel is \textit{Gaia} $G$ and 2MASS $K_S$, middle panel is 2MASS $J$ \& $K_S$, and the right panel is PS1 $r$ \& $z$. The PARSEC (version 1.2S) isochrone is for a metallicity $\mh=-0.2$ and age of 3.5~Gyr. The black dots are all stars found in each of the \textit{Gaia}, 2MASS, and Pan-STARRS1 catalogues within 7~arcmin of the cluster centre. Overplotted are the stars observed by HERMES (red dots) and AAOmega (blue squares) within 10~arcmin of the cluster. Errorbars show the $2\sigma$ error in magnitude and colour in 1~magnitude bins. These were the stars considered most likely to be members. The pink dots show the remaining stars that were observed across the entire 2dF field of view. For the HERMES observations these had no colour cuts, while for the AAOmega observations, these potential extratidal stars were restricted to the same colour and magnitude regions as the sub-giant branch and lower red giant branch observations.}
    \label{fig:gaia1_cmd_candidates}
\end{figure*}

For the HERMES observations, the highest priority targets were those within 5~arcmin of the cluster centre that photometry showed were in the prominent red clump (or red horizontal branch; 15 candidates observed), with the next highest priority given to the likely red giant branch (RGB) members (eight candidates). Lower priority targets were those in an annulus from 5--10~arcmin from the cluster and still photometrically located in either the clump or on the RGB (17 candidates). A further 287 stars were observed across the full 2~degree field within the magnitude range of $12<G<14$ and no colour cuts. The locations of the observed targets are shown on the colour-magnitude diagrams (CMD) in Figure \ref{fig:gaia1_cmd_candidates}. We also allocated 50 fibres to sky positions as it was unclear how bright the sky background would be due to the proximity of Sirius. The field was observed for three 20-minute exposures along with the standard exposures of the fibre flat lamps and the ThXe arc lamp. 

On the AAOmega nights, we concentrated on fainter stars within 7~arcmin of the cluster centre, with a selection of potential sub-giant branch (SGB) stars ($1.5<G-K_S<3.3; 17.4>G>16.5$;  42 candidates observed) and lower RGB stars ($1.5<G-K_S<2.3; 16.5>G>15.1$; 56 candidates). On the first night, 12 of the clump stars that had been identified as members from the HERMES spectra were also re-observed. We observed an additional 571 stars randomly chosen across the full two-degree field of view that were in the magnitude and colour cuts of the AAOmega RGB and SGB selections. These targets were divided into two fields, with one field observed on each night. The locations of the observed targets are shown on the CMD in Figure \ref{fig:gaia1_cmd_candidates}. The standard 25 sky fibres were used on both AAOmega nights, as the results from the HERMES observations confirmed that removing the sky background from Sirius was not going to be a problem. On the first night, we obtained three 20-minute exposures and on the second night four 20-minute exposures, along with the standard exposures of the fibre flat lamp and arc lamps. Unfortunately, it was found that the blue AAOmega spectra were dominated by scattered light from Sirius, and therefore could not be analyzed for this work.

\begin{figure}
    \includegraphics[width=\columnwidth]{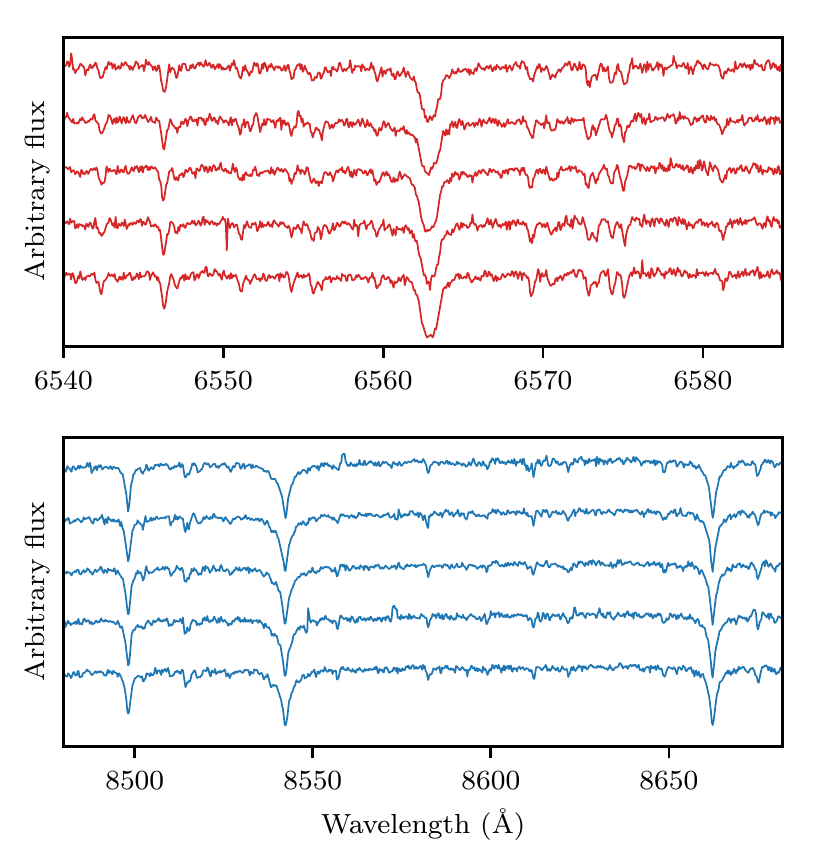}
    \caption{Examples of the reduced spectra for five clump stars. Top: A portion of the HERMES spectra centred on the H$\alpha$ line; Bottom: the same stars, showing their red AAOmega spectra and the calcium triplet lines. $\textrm{SNR}_\textrm{HERMES}\approx30$ per pixel and $\textrm{SNR}_\textrm{AAOmega}\approx50$, with the stars having $G$ magnitudes of about 14.8 (i.e., the faintest HERMES stars and the brightest AAOmega stars).}
    \label{fig:star_spectra_1}
\end{figure}

For both the HERMES and AAOmega observations, the raw spectra were reduced using the AAO's \textsc{2dfdr} reduction software \citep[version 6.28]{AAOSoftwareTeam2015} with the defaults for the particular spectrograph and gratings. Examples of the final reduced spectra for five red clump stars are shown in Figure \ref{fig:star_spectra_1}.

\section{Radial velocities}\label{sec:rv_feh}

For the stars observed with HERMES, the barycentric radial velocity was measured independently from the spectra of the blue, green, and red cameras\footnote{The near-infrared spectrum acquired by HERMES is partially dominated by terrestrial atmospheric bands, which make radial velocity measurements via cross-correlation difficult.} by cross-correlating the observed spectra with a template of the cool giant Arcturus. This was implemented with \textsc{iSpec}, an open source framework for spectral analysis \citep{Blanco-Cuaresma2014}. The average value obtained from the three cameras was adopted as the radial velocity of the star. 24 of the 327 stars observed returned inconsistent radial velocities between the three arms ($\sigma_V>10$~km/s). Inspection of their spectra and location on the colour-magnitude diagram revealed that they all tended to be the bluest stars observed ($G-K_S<1.5$) and therefore are likely hot dwarfs that are too dissimilar from Arcturus for cross-correlation to work successfully.

For the red AAOmega spectra, the near-infrared calcium triplet (CaT) lines at 8498.03, 8542.09 and 8662.14 \AA\ \citep{Edlen1956} were used to measure the barycentric radial velocities of the stars and to estimate their metallicity (for discussion of the metallicity results from the CaT lines, see Section \ref{sec:cat}). We direct the readers to \citet{Simpson2016f} for a full description of the method used to measure the radial velocities and equivalent widths of the CaT lines. Briefly, the spectra were normalized with a fifth-degree Chebyshev polynomial using \textsc{scipy}'s \textsc{chebfit} function and each CaT line fitted with a Voigt function provided by the \textsc{voigt1d} \citep{McLean1994} function from \textsc{astropy} \citep{Robitaille2013}. The central wavelength of the fitted Voigt functions was found and the average of the three line values adopted as the radial velocity of the star. This was repeated for 100 realizations with random Gaussian noise added to each wavelength pixel to understand the uncertainties of the method. The equivalent widths of the lines used for metallicity estimation (see Section \ref{sec:cat}) were found from the fitted Voigt functions.

\begin{figure}
    \includegraphics[width=\columnwidth]{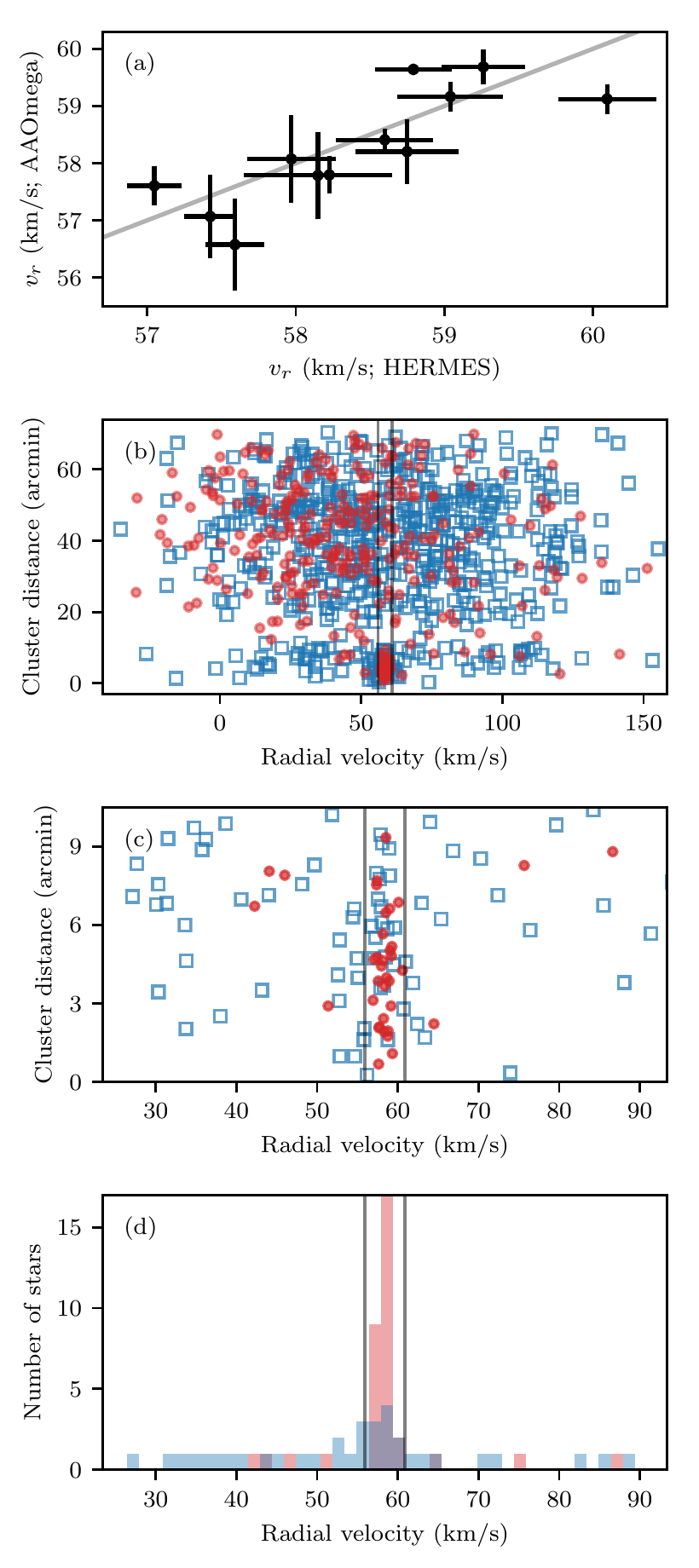}
    \caption{Radial velocity results. (a): Comparison of the radial velocity derived from the HERMES and AAOmega spectra for the stars in common. The line is the one-to-one line. (b): the radial velocities of all of the stars against their distance from the cluster centre of \gaia. Stars observed with HERMES are red dots and stars observed with AAOmega are blue squares. The lower density of targets observed in the range of 10--20~arcmin is due to the Sirius exclusion zone. (c): the same as (b) but showing only the inner 10~arcmin. (d): Histogram of the inner 10~arcmin, again with the HERMES velocity used for doubly observed stars. The vertical lines in the (b), (c) \& (d) indicate the velocities used to define the radial velocity range of probable cluster members.}
    \label{fig:radial_velocity}
\end{figure}

There were 12 stars observed with both HERMES and AAOmega. The top panel of Figure \ref{fig:radial_velocity} shows a comparison of the velocities derived from the two sets of observations; the consistency is on the $\lesssim 1$~km/s level. Inspection of the radial velocities of the candidates showed the strong velocity signature of the cluster. Of the 40 candidates observed by HERMES within 10~arcmin of the cluster, 29 stars had radial velocities with $56<r_v< 61$~km/s (Figure \ref{fig:radial_velocity}c). For those stars observed with AAOmega, in addition to the 12 stars in common with the HERMES observations, there were 21 probable members identified from their radial velocities using the velocity cut defined from the HERMES observations. Combining the results from the two datasets we find that the systemic radial velocity and dispersion of the cluster is $v_r=58.30\pm0.22$~km/s with a dispersion of $\sigma_v=0.94\pm0.15$~km/s (for just the HERMES sample $v_r = 58.30\pm0.10$~km/s and $\sigma_v=0.86\pm0.09$~km/s; and for just the AAOmega sample $v_r = 58.20\pm0.01$~km/s and $\sigma_v=1.0\pm0.01$~km/s).

\section{Metallicities}\label{sec:feh}
The metallicities of the probable cluster members identified in Section \ref{sec:rv_feh} were inferred from both the HERMES (Section \ref{sec:hermes_feh}) and AAOmega spectra (Section \ref{sec:cat}).

\subsection{HERMES spectra}\label{sec:hermes_feh}
For the stars observed with HERMES and identified as members in Section \ref{sec:rv_feh}, stellar parameters were determined from the high-resolution spectra with the classical method. The equivalent widths of the neutral iron lines were measured using \textsc{ares2} \citep{Sousa2015} and the ionized iron lines using \textsc{iraf}, and then the 1D LTE abundance for each line was calculated with \textsc{moog} \citep{Sneden1973} using Kurucz model atmospheres interpolated from the \citet{Castelli2004} grid of model atmospheres. The spectroscopic \teff\ was derived by requiring excitation equilibrium of Fe~I lines. The \logg\ was derived via ionization equilibrium, i.e., requiring the abundances from Fe~I lines to equal those from Fe~II lines. For comparison, we also computed photometric gravities via the Stefan-Boltzmann relation, using our spectroscopically derived \teff, bolometric corrections calculated from table 12 of \citet{Jordi2010}, a distance modulus of $(m-M)_G=14.50$, and with a stellar mass of $1.5$~M$_\textrm{\sun}$. Microturbulence was derived from the condition that abundances from Fe~I lines show no trend with equivalent width. For the 27 stars with reliable metallicity values, we find an average metallicity (and standard deviation of the sample) of $\feh=-0.13\pm0.13$. The results for each star are found in Table \ref{table:hermes_stars}. Uncertainties for the metallicities were found from the standard deviation of the Fe abundances found for each individual iron line divided by the square root of the number of iron lines used for each star.

While this paper was in preprint, \citet{Mucciarelli2017a} presented an analysis of six He-clump stars in \gaia\ based on spectra taken with the Magellan MIKE spectrograph. They note that the surface gravities we derive spectroscopically for some of our stars are 0.5 dex too high relative to an isochrone with age of 3 Gyr and solar metallicity (similar to our favored values of 3.5 Gyr and $\feh=-0.2$). Our HERMES spectra are lower signal than one would typically use for accurate spectroscopic abundance determination (due to limitations of telescope time available). However, as none of our primary conclusions depend on spectroscopic gravities, we do not believe this limitation will have had any impact on the nature, mass, age, stellar membership or orbit we find for \gaia,  or the metallicities we derive from the CaT region of the AAOmega spectra.  Indeed, within the uncertainties, our measurements of [Fe/H] agree with those derived by \citet{Mucciarelli2017a} from higher resolution spectra.

\subsection{AAOmega spectra}\label{sec:cat}

The metallicity of the stars observed with AAOmega was estimated from the strengths of their calcium triplet (CaT) lines. These lines have been used extensively in globular cluster studies to estimate the metallicity of member stars and there are a number of available empirical relationships that relate the metallicity of a star to its CaT line strengths and luminosity. In this work, as in \citet{Simpson2016f}, we have used \citet{Carrera2013}. This calibration has a valid metallicity range of $-4.0<\feh<+0.5$.

There are two key parameters for the CaT method: (1) the equivalent width measurements of the CaT lines (described in Section \ref{sec:rv_feh}); and (2) the luminosity of the star. As in \citet{Simpson2016f} we used the \citet{Carrera2013} empirical relationships with the absolute $K_S$ magnitude of the star. The absolute magnitude was found using the apparent magnitudes of the stars from the 2MASS catalogue and a distance modulus of $(m-M)_{K_S}=13.45\pm0.10$ (see Table \ref{table:extinction} and Section \ref{sec:parameters} for the determination of this distance modulus.).

\section{Cluster membership}\label{sec:members}

\begin{figure}
    \includegraphics[width=\columnwidth]{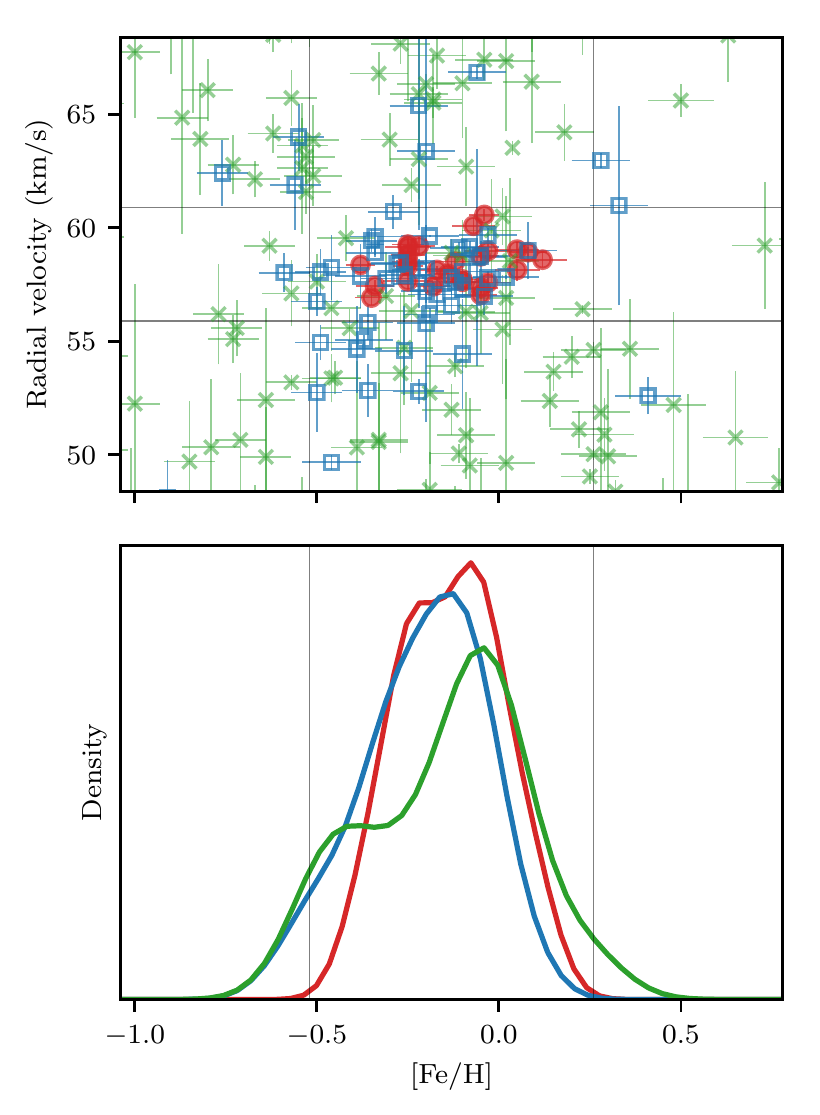}
    \caption{Top: The barycentric radial velocity of the stars against their metallicity. Red dots are stars observed with HERMES, blue squares are stars observed by AAOmega and within 10~arcmin of the cluster, and green crosses are stars observed by AAOmega and outside of 10~arcmin. The horizontal and vertical lines define the box for the membership selection. Bottom: Univariate kernel density estimate (with the same bandwidth of 0.5) of the metallicity of the different samples of members.}
    \label{fig:rv_feh}
\end{figure}

In Figure \ref{fig:rv_feh} the metallicity distributions of the two samples are plotted. There is good agreement between the metallicities found from the high- and low-resolution spectra. There is a larger spread of metallicities for stars observed with AAOmega than found for HERMES, so a metallicity cut was applied to the AAOmega results that is defined as the $3\sigma$ range of the HERMES metallicity results (the vertical lines in the top panel of Figure \ref{fig:rv_feh}). 

From the CaT method, for the 11 stars photometrically identified as clump members, and observed with AAOmega, we derive an overall metallicity of $\feh=-0.14\pm0.06$; for the 14 RGB members $\feh=-0.30\pm0.13$; and for the six SGB members $\feh=-0.09\pm0.13$. Overall, the metallicity estimate from the CaT method for the all 31 members observed with AAOmega was $\feh=-0.20\pm0.15$. As in \citet{Simpson2016f}, the CaT method is found to slightly underestimate the metallicity with respect to values derived from classical methods.

Most of the stars observed were further than 10~arcmin from the cluster, with the aim of identifying `extra-tidal' stars with radial velocities and metallicities matching those of cluster. From Figure \ref{fig:radial_velocity}(b), it is clear that the stars outside the tidal radius have a wide range of radial velocities, as would be expected for a random line of sight through the Galaxy \citep[e.g., see the results from RAVE survey;][]{Kunder2016}. In the brighter HERMES sample, all of these stars that were within the radial velocity limits for the cluster (defined in Section \ref{sec:rv_feh}) had bluer colours than the identified cluster members. Their positions on the CMD are not consistent with the cluster but with the field population of dwarf stars.

In the AAOmega sample, it is not as easy to exclude these large angular distance stars simply by considering their position on the CMD. The SGB and lower RGB of the cluster have the same brightness and colours as most of the field stars along the line of sight. These `extra-tidal' stars were also selected with colour and magnitude cuts designed to pick out stars on the expected RGB and SGB. In the top panel of Figure \ref{fig:rv_feh} we plot of the distribution of the metallicity versus the radial velocity of the stars within 10~arcmin and those stars observed by AAOmega that were further out (green crosses). Using the radial velocity and metallicity limits defined by the cluster sample, there are 15 potential ``extra-tidal'' stars of \gaia. Further observations of these stars will be required to determine if they are truly escaped members of the cluster.

\begin{figure}
    \includegraphics[width=\columnwidth]{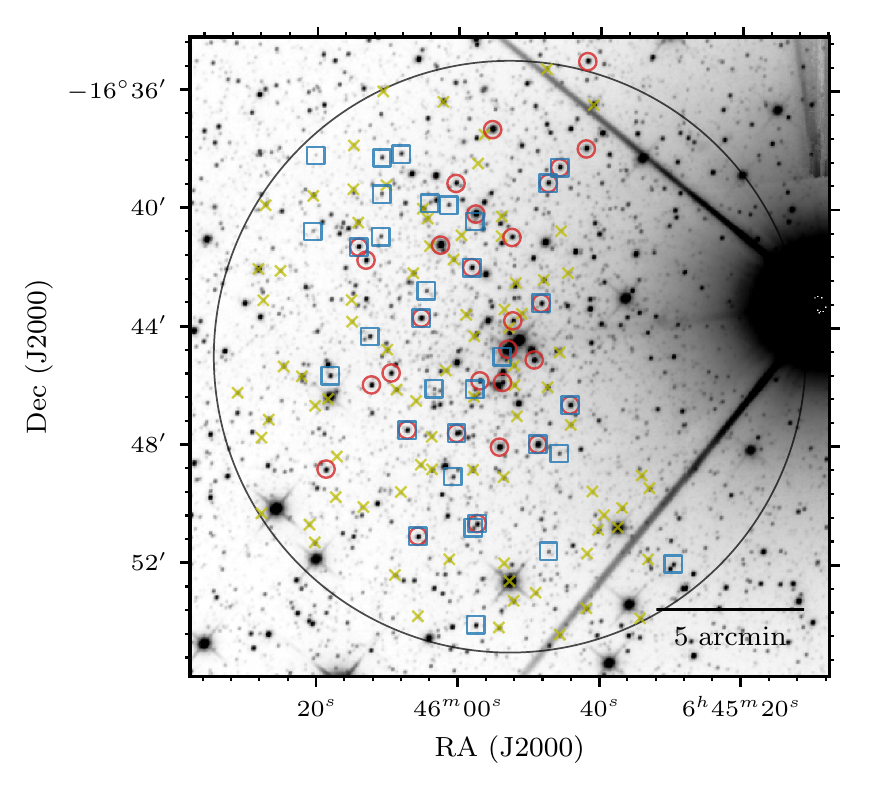}
    \caption{Sky distribution of the identified members (red circles: HERMES; blue squares: AAOmega) on the un-WISE W1 image \citep{Lang2014, Meisner2016}. Indicated with yellow crosses are field stars within 10~arcmin that were also observed. The large circle shows 10~arcmin around the cluster centre. The radial extent of the cluster is not clear as the exclusion zone around Sirius in our observing strategy (and also in 2MASS) has limited our ability to identify how far east the cluster extends on the sky.}
    \label{fig:unwise}
\end{figure}

The locations on the sky of the cluster members and field stars identified within 10~arcmin of the cluster centre are shown in Figure \ref{fig:unwise}. The radial extent of the cluster in the east-west direction on the sky is currently unclear. In the radial direction away from Sirius, it appears the edge of the cluster has been reached, as we do not find any radial velocity members further east than $\textrm{RA}=6^\textrm{h}46^\textrm{m}20^\textrm{s}$. In the radial direction towards Sirius, the edge is our observing exclusion zone around Sirius (see Section \ref{sec:observations}). This has placed an artificial limit on the identification of members radially towards Sirius.

\section{Discussion}\label{sec:parameters}

\begin{table}
\centering
\caption{Distance moduli and reddening determined by isochrone fitting (see Section \ref{sec:parameters}) from \textit{Gaia}, 2MASS, PS1 photometry. $(m-M)_0$ was determined for each using extinctions for different bandpasses determined by \citet{Schlafly2011}. They have not determined the values for \textit{Gaia} $G$ so no transformation was done.}
\label{table:extinction}
\begin{tabular}{rrrrrrr}
\hline
& & $(m-M)_X$ & $E(A-B)$ & $(m-M)_0$ & d (kpc)\\
\hline
$G$   & $G-K_S$ & $14.50$  & $0.80$   &  \\
$K_S$ & $J-K_S$ & $13.45$ & $0.27$   & 13.2 & 4.5 \\
$r$   & $r-z$   & $13.74$ & $0.53$   & 13.1 & 4.1 \\
\hline
\end{tabular}
\end{table}

\begin{figure*}
    \includegraphics[width=\textwidth]{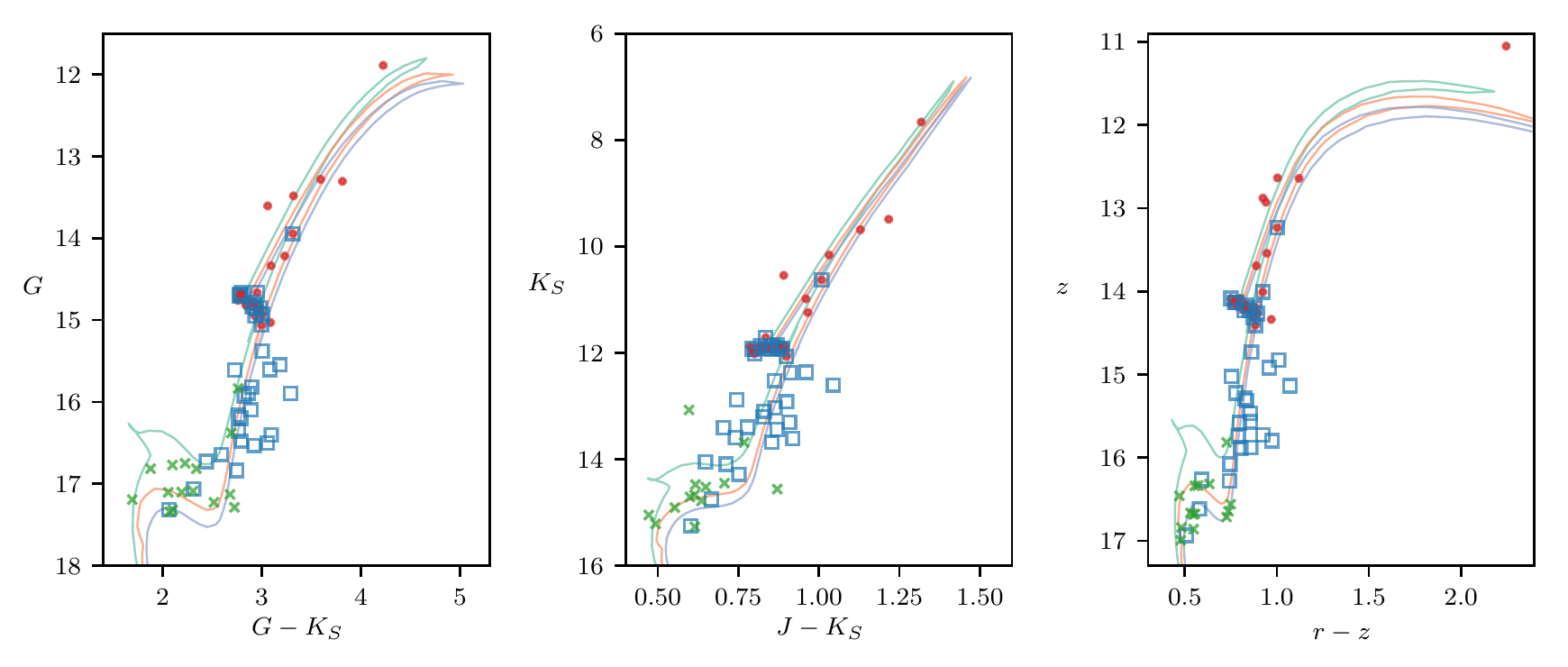}
    \caption{The CMDs in \textit{Gaia}, 2MASS and PS1 photometry (as in Figure \ref{fig:gaia1_cmd_candidates}, but showing only the probable members identified from their radial velocity and metallicity (red dots: observed with HERMES; blue squares: observed with AAOmega and $r<10$~arcmin; green cross: observed with AAOmega and $r>10$~arcmin). The isochrones are all for $\mh=-0.2$ and with ages of 2.5, 3.5, 4.5~Gyr (turn-off magnitude increases with age).}
    \label{fig:gaia1_cmd_members}
\end{figure*}

In Figure \ref{fig:gaia1_cmd_members} we replot the CMDs from Figure \ref{fig:gaia1_cmd_candidates}, with only the identified members and potential extratidal stars. The PARSEC isochrones \citep[version 1.2S;][]{Bressan2012,Chen2014,Chen2015,Tang2014} used have been primarily fitted by eye to the brightness and colour of the red clump stars. The luminosity of the red clump has only a weak dependence on the age \citep[e.g.,][]{Girardi2016} and so it was used to determine the distance modulus and reddening of the clusters (Table \ref{table:extinction}). The red clump brightness, the slope of the giant branch, and the turn-off magnitude are consistent with a $\mh=-0.2$ cluster with an age of about 3~Gyr.

With the aid of the spectroscopic results, we have fitted a more metal rich and younger isochrone than \citet{Koposov2017}, who used a representative $\feh=-0.7$, 6~Gyr isochrone. This has the result of making the tip of the giant branch fainter than their prediction, and as a consequence is that it seems very unlikely that there are any naked eye members of \gaia\ as suggested in the discovery paper. It is clear, however, that overall \gaia\ is bright enough that it would have been identified earlier either with the aided eye by astronomers in the 19th or 20th Century or during imaging surveys were it not hidden by the glare of Sirius.

There is a small discrepancy in the de-reddened distance moduli found using the optical and infrared photometry (Table \ref{table:extinction}). This is likely related to the transformations \citep[from][]{Schlafly2011} between the different photometric systems. For two reasons the distance modulus derived from the infrared photometry is preferred: (1) there is relatively high reddening in the direction of the cluster; and (2) the $K_S$ band has been shown to minimize the intrinsic differences in red clump star luminosities associated with metallicity \citep{Girardi2016}.

Using this distance modulus, the cluster is found at a heliocentric distance of $4.46\pm0.21$~kpc and a Galactocentric distance of $11.5\pm0.2$~kpc ($[X,Y,Z]=[11.0\pm0.1,-3.2\pm0.2,-0.64\pm0.03]$~kpc in a left-handed coordinate frame). The cluster half-light radius $6.5\pm0.4$~arcmin \citep{Koposov2017} translates to a physical half light radius of $8.4\pm0.6$~pc. These coordinate transformations were performed using \texttt{astropy}, assuming that the cluster position has an uncertainty of 1~arcmin and \texttt{astropy}'s default parameters that the Sun is 8.3~kpc from the Galactic centre and 27~pc above the plane \citep{Chen2001,Reid2004,Gillessen2009}.

\citet{Schlafly2011} and \citet{Green2015} estimated reddening values of $\ebv=0.4911\pm0.0079$ and $\ebv=0.36\pm0.031$ respectively for the direction and distance of \gaia. Transforming the reddening value found from the isochrone fit via relationships derived from \citet{Schlafly2011}, finds,
\begin{equation}
    \ebv=2.4\times\ejk=0.66.
\end{equation}
This reddening is twice what is predicted. This could be the result of the angular proximity to Sirius impacting the reddening estimates, or the accuracy of the photometry.

The metallicity we have found for \gaia\ is higher than the mean value for open clusters at its Galactocentric distance of 11.8~kpc \citep[e.g.,][]{Jacobson2009,Yong2012}, but it is within the distribution. The Galactic radial metallicity gradient as measured from open clusters has a transition at around $R_{GC}=13$~kpc \citep[e.g.,][]{Twarog1997}, becoming distinctly shallower in the outer disk. The similarity in the radial metallicity gradients for open clusters of different ages is commonly interpreted to mean that the Galactic metallicity gradient has been fairly stable over time \citep{Friel1993}, though possibly the transition radius has shifted outward \citep{Jacobson2011}. From its position and metallicity, \gaia\ would appear to belong to the inner disk population. However, our calculation of its orbit (Section \ref{sec:orbit}) would appear to contradict this explanation.

Assuming that \gaia\ is an isolated system for which the Virial Theorem holds, then the dynamical mass can be given by
\begin{equation}
    M_\mathrm{dyn} \simeq 2.5 \frac{3\sigma_0^2 r_h}{G} = 12.9^{+4.6}_{-3.8}\times10^3~\mathrm{M}_{\sun}.
\end{equation}
This value is very close to the 14000~M$_{\sun}$ estimated by \citet{Koposov2017} using the stellar density profile. The dynamical masses are dependent on the square of the velocity dispersion, and if the velocity dispersion is reduced from $\sigma_v = 0.94\pm0.15$~km/s (the overall value) to $0.86\pm0.09$~km/s (derived from just the HERMES sample), this reduces the cluster mass estimate by over 1000 solar masses.

\subsection{Orbit}\label{sec:orbit}

\begin{figure}
    \includegraphics[width=\columnwidth]{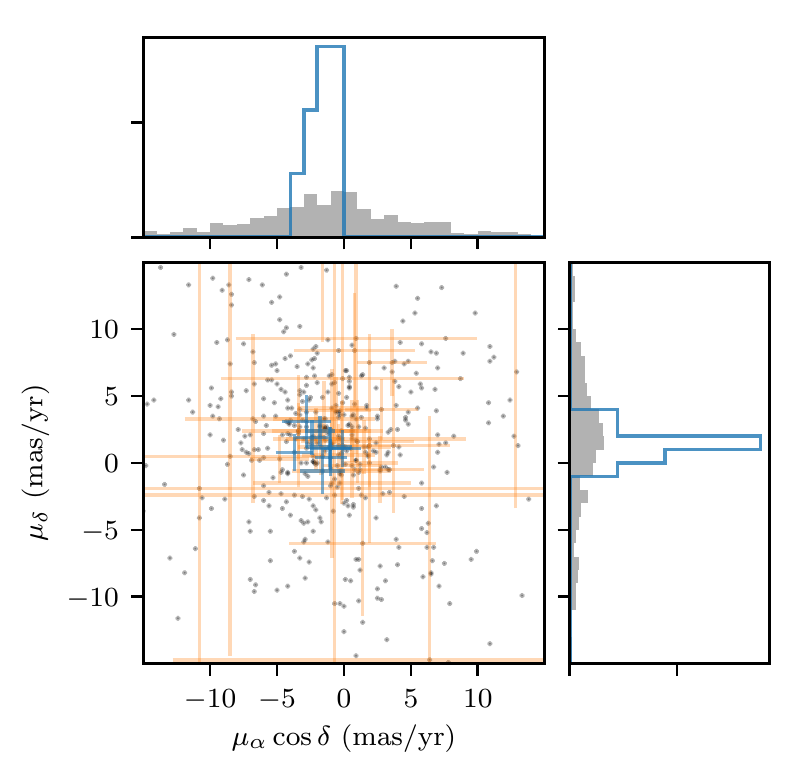}
    \caption{The UCAC5 proper motions of member stars with UCAC magnitudes less than 15 (blue error bars) and the rest of the identified members (orange crosses). The black dots are all UCAC5 targets within 12~arcmin of the cluster. The histograms show the distribution of the bright members (open blue) and all targets (shaded black histogram). A magnitude cut of 15 was selected based on figure 9 of \citet{Zacharias2017}, which shows the proper motion errors rapidly increasing at that brightness.}
    \label{fig:pm_plot}
\end{figure}

\begin{figure}
    \includegraphics[width=\columnwidth]{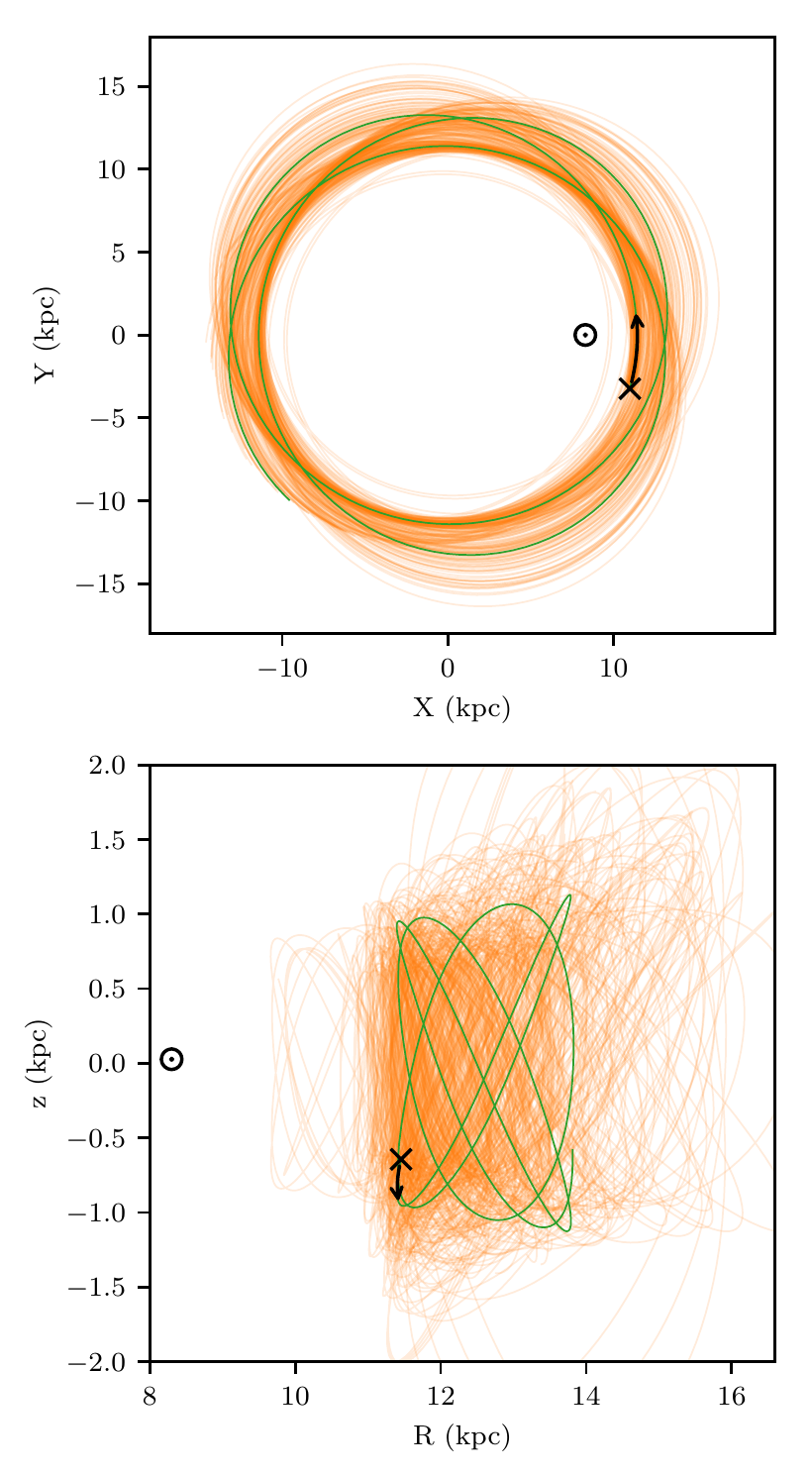}
    \caption{Projection of the orbit of \gaia\ integrated forward in time using \textsc{galpy}. The green line shows the orbit using the best values found for the cluster, and the fainter orange lines show the orbits of the random realizations. The black arrow indicates the direction of motion for this `best' orbit from the starting position. The currently observed position of \gaia\ is marked with a cross, and also shown is the present day of the Sun ($\sun$). For clarity, only 100 of the 10000 random realizations are shown.}
    \label{fig:orbit}
\end{figure}


The positional information was combined with kinematic information to estimate a probable orbit for the cluster. None of the stars identified as members were bright enough to be part of the Tycho-Gaia Astrometric Solution \citep[TGAS;][]{Michalik2014,Lindegren2016}, but 42 member stars were in the UCAC5 proper motion catalog \citep{Zacharias2017} and are shown on Figure \ref{fig:pm_plot}. A UCAC magnitude cut of $u_\textrm{mag}<15$ was selected based on figure 9 of \citet{Zacharias2017}, which shows the proper motion errors rapidly increasing at that brightness. There were nine stars that made this magnitude cut, all of which had small uncertainties in their proper motion: $\sigma(\mu_\alpha\cos\delta,\mu_\delta)<(1.8,2.0)$~mas/yr. Their mean proper motions were $(\mu_\alpha\cos\delta,\mu_\delta)=(-1.7\pm0.5,1.3\pm0.5)$~mas/yr.

We computed the orbits of the cluster using the \textsc{galpy} code \citep[\url{http://github.com/jobovy/galpy};][version 1.2]{Bovy2015} with inputs of $(\alpha,\delta,r_\textrm{\sun},\mu_\alpha\cos\delta,\mu_\delta,v_r)$ and the recommended simple Milky-Way-like \texttt{MWPotential2014} potential with the default parameters, and the Solar motion defined by \citet{Schönrich2010}. The cluster orbit was integrated forward in time for 1~Gyr with 1~Myr resolution, for 10000 random realizations varying the inputs with Gaussian errors. In Figure \ref{fig:orbit}, for clarity, a subset of 100 of these realizations are shown.

The median value of the orbital parameters was found for the 10000 realizations, with uncertainty ranges given by the 16th and 84th percentile values: the maximum and minimum Galactic distance achieved by the cluster are $r_\textrm{max}=13.8^{+1.4}_{-1.1}$~kpc and $r_\textrm{min}=11.4^{+0.2}_{-0.2}$~kpc; the largest distance out of the Galactic plane, $z_\textrm{max}=1.1^{+0.4}_{-0.3}$~kpc; and the eccentricity of the orbit $e=0.09^{+0.04}_{-0.03}$. The uncertainties in these orbital parameters are primarily driven by the uncertainty in the proper motions, with a much smaller contribution from the uncertainty in the distance. There was a negligible contribution from the uncertainty in the radial velocity and position of the cluster. The \textit{Gaia} DR2+ results should improve the precision to which the cluster's orbit can be calculated by providing accurate and precise proper motions.
 
The present day finds \gaia\ at about two-thirds of its maximum distance out of the Galactic plane. \citet{VandePutte2010} investigated the orbits of Galactic open clusters and found that most clusters are in quasi-periodic crown orbits like that of \gaia. They further classified clusters based upon their $z_\textrm{max}$ and radial quantity, defined as
\begin{equation}
    \eta = \frac{R_\textrm{max} - R_\textrm{min}}{0.5(R_\textrm{max} + R_\textrm{min})},
\end{equation}
where $R$ is the Galactocentric distance of the cluster projected onto the Galactic plane. For \gaia\ the radial quantity is $\eta=0.19^{+0.09}_{-0.06}$. It would be expected that clusters that formed in the disc of the Galaxy would have low $\eta$, and \citet{VandePutte2010} found 80 per cent of the 439 clusters had $\eta<0.28$, and 90 percent of clusters had $z_\textrm{max}<0.35$~kpc. Although \gaia\ has a circular orbit like the majority of open clusters, it has a large maximum distance out of the Galactic plane; over 3 times the scale height of the thin disk.

Numerical simulations of open cluster orbital evolution have shown that it is possible for the spiral arms of the Galaxy to have a large vertical effect on clusters, giving them large out-of-plane excursions (>200 pc), though these orbits tend to be chaotic \citep[see e.g.,][]{MartinezMedina2016}. \citet{MartinezMedina2017} investigated the survival of such high-altitude open clusters, and find that clusters in the plane of the disk and clusters with relatively large vertical motions ($z_\mathrm{max}\gtrapprox3.5$~kpc) tend to have the longest lifetimes. This is because the clusters experience the tidal stresses associated with disk crossings never (in the case of clusters in the plane) or rarely (in the case of high-altitude clusters). The clusters with the shortest lifetimes (with respect to an identical cluster on an in-plane orbit) have $z_\mathrm{max}\sim600$~pc. \gaia, with $z_\mathrm{max}\approx1.1$~kpc, is in a region of orbital parameter space which should be quite detrimental to its long-term survival. In its present orbit it makes nine plane crossings every gigayear, for a total of over 30 in its ~3 Gyr lifetime.

It is therefore surprising to find \gaia\ in its present orbit at the present day. This suggests that either it has recently moved into this orbit, perhaps after an interaction with a spiral arm \citep{MartinezMedina2016}, or that it has experienced significant mass loss in the past, and is now on the verge of complete destruction. Significant mass loss in the past would tend to support the association of the extratidal stars we find with radial velocities, metallicities and photometry consistent with \gaia. Higher precision radial velocity measurements could help to clarify whether \gaia\ is in virial equilibrium or whether it is in the process of disrupting.

\bigskip
We have presented the first spectroscopic observations of stellar cluster \gaia, which was recently discovered by \citet{Koposov2017}. Although initially these observations were carried out to investigate the novelty of a cluster that had previously been blocked from our view by glare from Sirius, these observations have shown \gaia\ is an interesting target in its own right, being relatively metal-rich and intermediate age cluster, and with a mass of $12.9^{+4.6}_{-3.8}\times10^3~\mathrm{M}_{\sun}$.

Both low and high-resolution spectra are consistent with the cluster having a metallicity of $\feh=-0.13\pm0.13$. Isochrone fits indicate that the cluster is about 3~Gyr in age. Orbital modelling shows that \gaia\ has a circular orbit but a large motion out of the plane of the Galaxy, and is currently found $640\pm30$~pc below the plane of the Galaxy and could travel as much as $z_\textrm{max}=1.1^{+0.4}_{-0.3}$~kpc. Such an orbit could result in the cluster experiencing over 30 plane passages during its lifetime, which means that \gaia\ could have a large stellar stream associated with it that is waiting to be discovered.

\section*{Acknowledgements}
JDS et al thank the other staff members at the AAO for their discussions about strategies for observing so close to Sirius. In particular Chris Lidman, who suggested placing Sirius at the centre of the field and using the black-coated 2dF plate. I am Sirius and don't call me Shirley.

We thank the anonymous referee for a useful report that improved the manuscript, and Chris Usher for a helpful correction to the original manuscript submitted to astro-ph.

Based on data acquired through the Australian Astronomical Observatory, under Director's Time for the HERMES observations, and under program A/2017A/102 for the AAOmega observations. The AAOmega observations were supported via the OPTICON H2020 Trans-National Access programme (Grant Reference Number 730890).

AMNF acknowledges support from the Shaw Visitor Scheme at the AAO, which enabled an extended visit during which this paper was completed. SLM and DBZ acknowledge support from Australian Research Council grants DE140100598 and FT110100743, respectively.

The following software and programming languages made this research possible: \textsc{configure} \citep{Miszalski2006}; the 2dF Data Reduction software \textsc{2dfdr} \citep[version 6.28;][]{AAOSoftwareTeam2015}; Python (version 3.5+); Astropy \citep[version 1.3;][]{Robitaille2013}, a community-developed core Python package for Astronomy; pandas \citep[version 0.19.2;][]{McKinney2010}; \textsc{topcat} \citep[version 4.3-5;][]{Taylor2005}; \textsc{galpy} \citep[version 1.2;][]{Bovy2015}. This research made use of APLpy, an open-source plotting package for Python \citep{2012ascl.soft08017R}.

This work has made use of data from the European Space Agency (ESA) mission {\it Gaia} (\url{http://www.cosmos.esa.int/gaia}), processed by the {\it Gaia} Data Processing and Analysis Consortium (DPAC, \url{http://www.cosmos.esa.int/web/gaia/dpac/consortium}). Funding for the DPAC has been provided by national institutions, in particular the institutions participating in the {\it Gaia} Multilateral Agreement.

The Pan-STARRS1 Surveys (PS1) and the PS1 public science archive have been made possible through contributions by the Institute for Astronomy, the University of Hawaii, the Pan-STARRS Project Office, the Max-Planck Society and its participating institutes, the Max Planck Institute for Astronomy, Heidelberg and the Max Planck Institute for Extraterrestrial Physics, Garching, The Johns Hopkins University, Durham University, the University of Edinburgh, the Queen's University Belfast, the Harvard-Smithsonian Center for Astrophysics, the Las Cumbres Observatory Global Telescope Network Incorporated, the National Central University of Taiwan, the Space Telescope Science Institute, the National Aeronautics and Space Administration under Grant No. NNX08AR22G issued through the Planetary Science Division of the NASA Science Mission Directorate, the National Science Foundation Grant No. AST-1238877, the University of Maryland, Eotvos Lorand University (ELTE), the Los Alamos National Laboratory, and the Gordon and Betty Moore Foundation.

This publication makes use of data products from the Two Micron All Sky Survey, which is a joint project of the University of Massachusetts and the Infrared Processing and Analysis Center/California Institute of Technology, funded by the National Aeronautics and Space Administration and the National Science Foundation.




\appendix
\newpage
\section{Line list}
\begin{table}
\centering
\caption{The line list used for the HERMES stellar parameter determination.}\label{table:line_list}
\begin{tabular}{rlrr}
\hline
$\lambda$ (\AA) & Element & EP (eV) & $\log gf$\\
\hline
4788.757 & FeI & 3.237 & $-1.763$\\
4794.360 & FeI & 2.424 & $-3.950$\\
4802.880 & FeI & 3.642 & $-1.510$\\
4808.148 & FeI & 3.251 & $-2.690$\\
4834.517 & FeI & 2.420 & $-3.330$\\
4890.755 & FeI & 2.875 & $-0.394$\\
4891.492 & FeI & 2.851 & $-0.111$\\
5651.469 & FeI & 4.473 & $-1.900$\\
5652.318 & FeI & 4.260 & $-1.850$\\
5661.346 & FeI & 4.284 & $-1.756$\\
5679.023 & FeI & 4.652 & $-0.820$\\
5696.090 & FeI & 4.548 & $-1.720$\\
5701.557 & FeI & 2.560 & $-2.160$\\
5704.733 & FeI & 5.033 & $-1.409$\\
5705.465 & FeI & 4.301 & $-1.355$\\
5720.898 & FeI & 4.548 & $-1.631$\\
5731.762 & FeI & 4.256 & $-1.200$\\
5732.296 & FeI & 4.991 & $-1.460$\\
5741.848 & FeI & 4.256 & $-1.672$\\
5752.032 & FeI & 4.549 & $-1.177$\\
5775.081 & FeI & 4.220 & $-1.297$\\
5778.453 & FeI & 2.588 & $-3.430$\\
5806.724 & FeI & 4.607 & $-1.030$\\
5809.218 & FeI & 3.883 & $-1.790$\\
5853.148 & FeI & 1.485 & $-5.180$\\
5855.077 & FeI & 4.608 & $-1.478$\\
5858.778 & FeI & 4.220 & $-2.160$\\
5861.110 & FeI & 4.283 & $-2.304$\\
5862.357 & FeI & 4.549 & $-0.127$\\
6494.994 & FeI & 2.400 & $-1.256$\\
6498.939 & FeI & 0.958 & $-4.688$\\
6546.239 & FeI & 2.758 & $-1.536$\\
6592.914 & FeI & 2.727 & $-1.473$\\
6593.870 & FeI & 2.433 & $-2.420$\\
6597.561 & FeI & 4.795 & $-0.970$\\
6627.545 & FeI & 4.548 & $-1.580$\\
6653.853 & FeI & 4.154 & $-2.215$\\
6677.987 & FeI & 2.692 & $-1.418$\\
6699.142 & FeI & 4.593 & $-2.101$\\
6703.567 & FeI & 2.758 & $-3.060$\\
6705.101 & FeI & 4.607 & $-1.392$\\
6710.319 & FeI & 1.485 & $-4.764$\\
6713.745 & FeI & 4.795 & $-1.500$\\
6725.357 & FeI & 4.103 & $-2.013$\\
6726.667 & FeI & 4.607 & $-1.133$\\
6733.151 & FeI & 4.638 & $-1.480$\\
7710.364 & FeI & 4.220 & $-1.113$\\
7723.210 & FeI & 2.280 & $-3.617$\\
7748.269 & FeI & 2.949 & $-1.751$\\
7751.109 & FeI & 4.991 & $-0.783$\\
7780.556 & FeI & 4.473 & $-0.010$\\
7802.473 & FeI & 5.086 & $-1.417$\\
7807.909 & FeI & 4.991 & $-0.521$\\
7844.559 & FeI & 4.835 & $-1.759$\\
4720.150 & FeII & 3.197 & $-4.822$\\
4731.453 & FeII & 2.891 & $-3.127$\\
4833.197 & FeII & 2.657 & $-4.795$\\
6516.080 & FeII & 2.891 & $-3.432$\\
7711.723 & FeII & 3.903 & $-2.683$\\
7841.390 & FeII & 3.900 & $-3.896$\\
\hline
\end{tabular}
\end{table}

\newpage
\section{Cluster members}
\begin{table*}
\centering
\caption{Parameters for members of \gaia\ observed with HERMES.}\label{table:hermes_stars}
\begin{tabular}{llrrrrrrrrrrr}
\hline
\textit{Gaia} source\_id & PS1 objID & $r$ ($\prime$) & SNR1 & SNR2 & SNR3 & SNR4 & $v_r$ (km/s) & \teff\ (K) & $\logg_{\mathrm{spec}}$ & $\logg_{\mathrm{phot}}$ & \feh \\
\hline
2946299839381625344 & 87771015236028832 & 7.70 & 10 & 19 & 32 & 32 & $57.4\pm0.3$ & 5200 & 3.5 & 2.7 & $-0.18\pm0.05$ \\
2946300011181433856 & 87781014887617472 & 6.86 & 8 & 15 & 25 & 25 & $60.1\pm0.6$ & 5400 & 4.0 & 2.8 & $-0.07\pm0.06$ \\
2946300423497294336 & 87821015778053968 & 7.53 & 10 & 19 & 30 & 30 & $57.4\pm0.4$ & 5400 & 3.5 & 2.7 & $-0.05\pm0.05$ \\
2946300664016492032 & 87851015301440560 & 4.68 & 11 & 20 & 33 & 32 & $57.0\pm0.3$ & 5300 & 3.6 & 2.7 & $-0.05\pm0.05$ \\
2946301041973608960 & 87831014757969200 & 4.26 & 15 & 31 & 58 & 57 & $60.6\pm0.4$ & 4900 & 3.1 & 2.1 & $-0.04\pm0.04$ \\
2946301076335192064 & 87841015010738448 & 3.98 & 9 & 17 & 29 & 26 & $58.6\pm0.6$ & 5600 & 4.3 & 2.9 & $0.12\pm0.07$ \\
2946301145053017600 & 87841014533031232 & 4.43 & 14 & 28 & 49 & 50 & $58.0\pm0.5$ & 4700 & 2.5 & 2.1 & $-0.14\pm0.04$ \\
2946301557369703424 & 87881015512560992 & 4.63 & 11 & 20 & 44 & 31 & $58.1\pm0.3$ & 5400 & 3.7 & 2.8 & $-0.17\pm0.06$ \\
2946301557369708544 & 87881015396809088 & 3.86 & 10 & 17 & 31 & 31 & $58.9\pm0.4$ & 5150 & 3.1 & 2.7 & $0.08\pm0.05$ \\
2946301729168396800 & 87881014741582944 & 2.10 & 13 & 22 & 33 & 35 & $57.7\pm0.6$ & 5450 & 3.6 & 2.7 & $-0.10\pm0.04$ \\
2946301729168397824 & 87881014874583952 & 2.07 & 8 & 14 & 29 & 28 & $57.6\pm0.6$ & 5300 & 3.5 & 2.7 & $-0.25\pm0.07$ \\
2946303378435861504 & 87921015220436288 & 2.42 & 9 & 18 & 29 & 31 & $58.2\pm0.4$ &  &  &  &  \\
2946303550234568704 & 87961015546585312 & 4.75 & 8 & 15 & 25 & 26 & $57.4\pm0.4$ & 5300 & 3.5 & 2.8 & $-0.34\pm0.05$ \\
2946303550234570752 & 87971015589864016 & 5.17 & 8 & 15 & 29 & 28 & $59.3\pm0.5$ & 5250 & 3.5 & 2.7 & $-0.25\pm0.04$ \\
2947051836616382464 & 87861014343327792 & 3.86 & 10 & 17 & 29 & 28 & $57.6\pm0.3$ & 5400 & 4.7 & 2.8 & $-0.03\pm0.06$ \\
2947052042774817792 & 87891014555928208 & 1.92 & 16 & 29 & 57 & 49 & $58.3\pm0.4$ & 4950 & 2.9 & 2.3 & $-0.25\pm0.04$ \\
2947052352012464128 & 87901014709095408 & 1.08 & 21 & 54 & -89 & 137 & $59.3\pm0.5$ &  &  &  &  \\
2947052455091682304 & 87921014682294608 & 0.68 & 15 & 24 & 36 & 36 & $57.6\pm0.5$ & 5300 & 3.0 & 2.7 & $-0.10\pm0.05$ \\
2947052729969764864 & 87931014515586560 & 1.74 & 8 & 14 & 29 & 27 & $58.7\pm0.6$ & 5400 & 3.7 & 2.8 & $-0.05\pm0.05$ \\
2947054276155475968 & 87961014925020480 & 1.94 & 14 & 24 & 34 & 36 & $58.8\pm0.4$ & 5100 & 2.5 & 2.6 & $-0.25\pm0.05$ \\
2947054310515227648 & 87971015108695600 & 3.12 & 14 & 34 & 540 & 73 & $56.9\pm0.5$ & 4900 & 2.8 & 2.0 & $-0.35\pm0.05$ \\
2947054379234722304 & 87991014902006688 & 3.66 & 16 & 35 & 88 & 72 & $58.4\pm0.8$ & 4450 & 2.2 & 1.7 & $-0.38\pm0.04$ \\
2947055100791549952 & 87981014689160944 & 2.90 & 12 & 19 & 38 & 34 & $59.2\pm0.6$ & 5400 & 3.2 & 2.7 & $-0.25\pm0.06$ \\
2947055272587941888 & 88011015019007344 & 4.82 & 10 & 17 & 32 & 31 & $59.2\pm0.7$ & 5100 & 3.0 & 2.6 & $-0.22\pm0.04$ \\
2947055375669662720 & 88011014476597600 & 5.02 & 13 & 21 & 35 & 35 & $59.0\pm0.6$ & 5200 & 2.7 & 2.6 & $0.05\pm0.06$ \\
2947055444388945408 & 88021014409238176 & 5.66 & 9 & 16 & 35 & 31 & $58.1\pm0.9$ & 6000 & 4.4 & 2.9 & $0.05\pm0.06$ \\
2947055547465894400 & 88051014804973792 & 6.47 & 16 & 33 & 67 & 63 & $58.5\pm0.6$ & 4500 & 2.5 & 1.8 & $-0.25\pm0.04$ \\
2947058399326636032 & 88041014253730976 & 6.62 & 14 & 25 & 46 & 43 & $59.0\pm0.6$ & 4950 & 3.5 & 2.4 & $-0.03\pm0.07$ \\
2947058914722539520 & 88101014246980016 & 9.34 & 13 & 22 & 39 & 38 & $58.5\pm0.5$ & 5200 & 3.2 & 2.6 & $-0.12\pm0.07$ \\
\hline
\end{tabular}
\end{table*}

\begin{table*}
\centering
\caption{Parameters for members of \gaia\ observed with AAOmega.}\label{table:aaomega_stars}
\begin{tabular}{llrrrrrrrrrr}
\hline
\textit{Gaia} source\_id & PS1 objID & $r$ (arcmin) & SNR & $v_r$ (km/s) & \feh \\
\hline
2946293723347668992 & 87711014894919104 & 9.13 & 71 & $58.0\pm1.7$ & $-0.49\pm0.07$ \\
2946294719781718016 & 87761013733870656 & 8.92 & 58 & $58.9\pm0.3$ & $-0.34\pm0.08$ \\
2946294822860913664 & 87761014467728992 & 6.70 & 32 & $57.9\pm0.1$ & $-0.26\pm0.07$ \\
2946299839381625344 & 87771015236028832 & 6.81 & 52 & $57.1\pm1.3$ & $-0.17\pm0.08$ \\
2946300011181433856 & 87781014910484720 & 5.91 & 36 & $59.6\pm1.5$ & $-0.34\pm0.07$ \\
2946300011181433856 & 87781014887617472 & 5.75 & 52 & $59.1\pm0.5$ & $-0.11\pm0.08$ \\
2946300217339877888 & 87811015031489296 & 4.48 & 51 & $59.4\pm0.9$ & $-0.35\pm0.07$ \\
2946300664016492032 & 87851015301440560 & 4.26 & 49 & $57.6\pm0.6$ & $-0.12\pm0.08$ \\
2946301076335192064 & 87841015010738448 & 3.15 & 46 & $58.4\pm0.3$ & $-0.28\pm0.08$ \\
2946301145053017600 & 87841014533031232 & 3.11 & 65 & $58.1\pm1.3$ & $-0.10\pm0.10$ \\
2946301179412755456 & 87831014407495008 & 3.67 & 47 & $58.3\pm2.5$ & $-0.46\pm0.07$ \\
2946301488650226176 & 87871015144668416 & 2.78 & 32 & $60.7\pm1.3$ & $-0.29\pm0.07$ \\
2946301660448917504 & 87871014904248352 & 1.61 & 26 & $58.8\pm1.2$ & $-0.05\pm0.08$ \\
2946303103557947392 & 87911015523073776 & 4.78 & 31 & $58.5\pm1.2$ & $-0.27\pm0.08$ \\
2946303550234570752 & 87971015589864016 & 6.30 & 40 & $59.7\pm0.5$ & $-0.03\pm0.08$ \\
2946303893831958016 & 87981015859664576 & 7.88 & 21 & $59.0\pm2.2$ & $0.08\pm0.08$ \\
2947051836616382464 & 87861014343327792 & 2.62 & 52 & $56.6\pm1.4$ & $-0.13\pm0.08$ \\
2947052352012463104 & 87901014745180368 & 0.26 & 34 & $56.2\pm0.5$ & $-0.19\pm0.07$ \\
2947052729969764864 & 87931014515586560 & 2.10 & 48 & $58.2\pm1.0$ & $-0.20\pm0.08$ \\
2947054001280092672 & 87941015191724624 & 3.59 & 34 & $57.9\pm2.4$ & $-0.38\pm0.07$ \\
2947054173076278784 & 87981015460201088 & 5.95 & 33 & $56.8\pm1.1$ & $-0.50\pm0.07$ \\
2947054276155475968 & 87961014925020480 & 3.27 & 42 & $59.6\pm0.1$ & $-0.19\pm0.08$ \\
2947054379234717184 & 87991014908411616 & 4.72 & 27 & $57.0\pm1.4$ & $-0.04\pm0.08$ \\
2947054452254690304 & 88001015172583952 & 5.85 & 25 & $58.7\pm8.3$ & $-0.06\pm0.08$ \\
2947054482313943040 & 88001015061852496 & 5.51 & 26 & $57.2\pm0.9$ & $-0.20\pm0.07$ \\
2947054619752901632 & 88001015455799920 & 6.99 & 20 & $57.5\pm1.9$ & $-0.22\pm0.07$ \\
2947054791551624704 & 88031015844675936 & 9.45 & 17 & $57.8\pm0.2$ & $0.02\pm0.09$ \\
2947054860271100928 & 88031015454464384 & 7.98 & 30 & $57.3\pm0.7$ & $-0.10\pm0.08$ \\
2947054928990581760 & 88031015341416976 & 7.76 & 53 & $57.8\pm1.2$ & $-0.31\pm0.07$ \\
2947055375669662720 & 88011014476597600 & 6.00 & 45 & $59.2\pm0.5$ & $-0.08\pm0.08$ \\
2947055444388945408 & 88021014409238176 & 6.61 & 49 & $57.8\pm1.3$ & $-0.13\pm0.08$ \\
\hline
\end{tabular}
\end{table*}

\bsp    
\label{lastpage}
\end{document}